\documentclass[twocolumn,eqsecnum,preprintnumbers,nofootinbib]{revtex4}
\usepackage{amsmath,amssymb,amsfonts,times,graphicx,color}
\pdfoutput=1

\newcommand{\beq}{\begin{equation}}
\newcommand{\eeq}{\end{equation}}
\def\bea{\begin{eqnarray}}
\def\eea{\end{eqnarray}}

\begin{document}

\title{Entanglement does not generally decrease under renormalization}
\author{Brian Swingle}
\affiliation{Department of Physics, Harvard University, Cambridge MA 02138}

\date{\today}
\begin{abstract}
Renormalization is often described as the removal or ``integrating out" of high energy degrees of freedom.  In the context of quantum matter, one might suspect that quantum entanglement provides a sharp way to characterize such a loss of degrees of freedom.  Indeed, for quantum many-body systems with Lorentz invariance, such entanglement monotones have been proven to exist in one, two, and three spatial dimensions.  In each dimension d, a certain term in the entanglement entropy of a d-ball decreases along renormalization group (RG) flows.  Given that most quantum many-body systems available in the laboratory are not Lorentz invariant, it is important to generalize these results if possible.  In this work we demonstrate the impossibility of a wide variety of such generalizations.  We do this by exhibiting a series of counterexamples with understood renormalization group flows which violate entanglement RG monotonicity.  We discuss bosons at finite density, fermions at finite density, and majorization in Lorentz invariant theories, among other results.
\end{abstract}

\maketitle

\section{Introduction}
Renormalization is often described as the removal or ``integrating out" of high energy degrees of freedom  \cite{RevModPhys.47.773}.  As such, one might imagine that entropic-like quantities measuring information would provide a sharp way to formulate the loss of information inherent in the renormalization process.  Entanglement is a natural target for investigation because it is known to be a precious resource in other contexts, for example, entanglement cannot be created between two separated parties using only local operations and classical communication (LOCC) \cite{PhysRevLett.83.436}.  Within the context of quantum ground states, a particularly natural quantity measuring entanglement is the entanglement entropy. This entropy measures how entangled a given spatial subregion is with its environment.  Because it often increases with the number of degrees of freedom, e.g., the number of fermion flavors, the entanglement entropy, or some generalization or refinement thereof, might provide a suitable way to quantify the loss of information along renormalization group (RG) flows from one quantum ground state to another.  Precisely such a result has been obtained for Lorentz invariant field theories in one \cite{Zc,2007JPhA...40.7031C}, two \cite{MSF1,MSF2,2011JHEP...06..102J,2012PhRvD..85l5016C}, and three spatial dimensions \cite{KSa,Cardy1988749}.  It is the purpose of this paper to show that such a result, despite its appeal, cannot generally be extended beyond the class of Lorentz invariant systems.

To understand these claims, we first introduce the concept of entanglement entropy.  Throughout the paper we consider quantum ground states of local Hamiltonians in $d$ spatial dimensions.  Let $|G\rangle $ be the ground state of such a local Hamiltonian.  Given a spatial subregion $A$ of the total many-body system, we may partition the Hilbert space as $\mathcal{H} = \mathcal{H}_A \otimes \mathcal{H}_B$ where $B$ is the complement of $A$ and $\mathcal{H}_R$ is the Hilbert space built from degrees of freedom in region $R$.  The state of the full system is \beq
\rho_{AB} = | G \rangle \langle G |
\eeq
and the state of subregion $A$ is
\beq
\rho_A = \text{tr}_B(\rho_{AB}).
\eeq
The entanglement entropy of $A$ in the global pure state $|G \rangle $ is then
\beq
S(A) = -\text{tr}_A(\rho_A \ln{\rho_A}).
\eeq
If $S(A)=0$ then $A$ and $B$ are not entangled and we have $|G\rangle = | G_A \rangle \otimes |G_B \rangle$ as in mean field theory.  More generally, $S(A)$ measures the degree to which $A$ is entangled with $B$.  For example, if $A$ and $B$ share $N_s$ singlets then $S(A)=S(B) = N_s \ln{(2)}$.

Now we are ready to understand the RG monotonicity results for Lorentz invariant field theories.  First off, we must regulate such a theory to render the entanglement finite.  A lattice regularization provides a simple method to render the theory finite, but we will not need to specify the regulator for our considerations here.  Indeed, if we consider the special subclass of conformal field theories (theories with a larger symmetry including scaling symmetry) then the details of the regulator should be irrelevant in the RG sense.\footnote{One could worry that the presence of Lorentz-breaking irrelevant operators, however small, invalidates the proofs of entanglement monotonicity.  However, because the entanglement monotone has a thermodynamic interpretation, it seems that the conclusion is robust.  At least no counterexamples are known in lattice regulated models.}  If $\epsilon$ is an ultraviolet cutoff length scale, then the entanglement entropy of a $d$-ball of radius $L$ in such a conformal field theory (CFT) has the form
\beq\label{Sc}
S_{d=1}(L) = \frac{c}{3}\ln{(L/\epsilon)},
\eeq
\beq \label{SF}
S_{d=2}(L) =  k_1 \frac{L}{\epsilon} - F,
\eeq
\beq \label{Sa}
S_{d=3}(L) = k_2 \frac{L^2}{\epsilon^2} - a \ln{(L/\epsilon)}.
\eeq
The constants $k_i$ are non-universal while the constants $c,F,a$ are universal (the nomenclature is conventional).  The main result of previous works is that $c$, $F$, and $a$ all decrease along the RG flow.  More precisely, if a UV CFT is perturbed by a relevant operator and flows to an IR CFT while preserving Lorentz invariance then the appropriate (depending on the dimension) entanglement quantity satisfies $Q_{UV} > Q_{IR}$ where $Q=c,F,a$ in $d=1,2,3$.

Thus we see that for the special class of Lorentz invariant RG fixed points, that is conformal field theories, entanglement entropy for a special subregion (a ball) does provide a sharp way to characterize the loss of information along the RG flow.  Such results are deep and deserve to stand on their own; for some recent early uses of such results within the theory of quantum matter see Refs. \cite{2012PhRvB..86o5131S,2012arXiv1211.1392G}.  Ref. \cite{2012PhRvB..86o5131S} made the argument that certain ``deconfined" critical points \cite{Senthil05032004} should be viewed as more highly entangled than conventional critical points, a fact that would have implications for the use of tensor network states \cite{mera,peps,terg} and DMRG \cite{dmrg_orig} to describe such deconfined critical points.  To justify this intuition, it is useful to appeal to an interpretation of the entanglement entropy of a $d$-ball, $S_d$, as the thermal entropy of the CFT on a cutoff version of hyperbolic space (the cutoff renders the otherwise divergent spatial volume finite) \cite{2011JHEP...05..036C}.  Neglecting surface contributions from the cutoff, the constant $k_2$, say, is then itself proportional to $F$, thus larger $F$ does imply larger area law term within this class of regulators.  Independently, Ref. \cite{2012arXiv1211.1392G} considered dynamical questions and showed that entanglement monotonicity, along with other plausible assumptions, constrains the phase structure of certain systems of gauge fields coupled to fermions of interest to condensed matter physics. In that context, for example, one shows that a system cannot break symmetry because the resulting low energy theory would have too many massless Goldstone modes and hence too large a value of the entanglement monotone.

Our motivation is thus two-fold: we would like to answer the fundamental question of whether entanglement is in some sense monotonic under RG flows and if it is, we would like to find new applications of entanglement monotonicity that illuminate the physics of quantum matter.  In this work we test whether entanglement monotonicity can be generalized to non-Lorentz invariant settings.  This is an important question since Lorentz invariance is absent in most forms of quantum matter, e.g., solid state systems, cold atoms, and finite density quark matter systems.  Unfortunately, such a generalization, if it exists, must be subtle.  For example, there are many measures of the number of degrees per unit volume, yet only $c$, $F$, and $a$ are monotonic.  The coefficient of the low temperature thermal entropy is not monotonic (see the discussion of majorization below), nor is the entanglement entropy for subregions besides $d$-balls.  More generally, the structure of the entanglement entropy, e.g., its dependence on region size $L$, is not even the same across non-Lorentz invariant theories, so what then should we compare between such very different systems?

Nevertheless, we can forge ahead and try to formulate a generalized monotonicity theorem.  There are certain basic requirements that should be obeyed by any proposed RG monotone.  After all, trivial quantities exist which always increase under renormalization, e.g., the RG time or the inverse cutoff.  However, these quantities are not properties of fixed points because they depend upon the whole RG flow.  A good monotone is something that is a property of the continuum limit of the theory and is independent of the microscopic regulator.  In a similar vein, it is not enough to say that the total number of degrees of freedom is decreasing, for example, when we integrate out high momentum modes in a Wilsonian RG \cite{RevModPhys.47.773}.  Instead, we see from the results above that it is certain ``intensive" measures of the number of degrees that are monotonic.  It is more like the number of degrees of freedom per unit volume which is decreasing, but only as measured by a certain special physical probe (the entanglement entropy of a ball in the ground state).  This is the sort of generalization which we seek.

In this paper we show that there is no simple generalization of the $c$, $F$, or $a$ theorem once Lorentz invariance is lost.  We do this by exhibiting a variety of simple, almost trivial, counterexamples that rule out many possible extensions.  Indeed, we will show that no quantity built purely from the ground state can decrease in general under an RG flow.  We will also show that some attempts to generalize the monotonicity theorems while staying within the class of Lorentz invariant systems also fail.  We give some simple fermionic examples in one dimension as a warm up.  Then we exhibit a simple RG flow for non-relativistic bosons from a theory with dynamical exponent $z=2$ to a theory with dynamical exponent $z=1$ which goes from being unentangled to being entangled like a Lorentz invariant conformal field theory.  Next we use the concept of Fermi surfaces to badly violate monotonicity in higher dimensions.  In the context of Lorentz invariant theories, we show that the Renyi entropy $S_n$ (a generalization of $S$ defined below) is not monotonic at small Renyi parameter $n \ll 1$ even for a $d$-ball when $d>1$.  Finally, we discuss the possibility of monotonicity theorems under restricted conditions, e.g. enhanced symmetry, and the overall outlook of our work.  The main message seems to be that the varieties of entanglement are greatly multiplied once Lorentz invariance is lost.

\section{Free fermions}

As a first example, we consider a fermionic hopping Hamiltonian on a one dimensional chain of $N$ sites.  We have fermions $c_r$ on each site of the chain with Hamiltonian
\beq
H = - \sum_r \sum_{\ell} w_\ell c^\dagger_{r+\ell} c_r + h.c.
\eeq
and periodic boundary conditions.  The $w_\ell$ are hopping matrix elements.  In terms of momentum space fermion operators
\beq
c_k = \sum_r \frac{e^{-i k r}}{\sqrt{N}} c_r
\eeq
the Hamiltonian is
\beq
H_c = \sum_{k} \epsilon_k c^\dagger_k c_k
\eeq
with
\beq
\epsilon_k = \sum_\ell -2 w_\ell \cos{(\ell k)}.
\eeq

$w_{\ell >1} = 0$ is the usual tight binding model with $\epsilon_k = - 2 w_1 \cos{k}$.  The ground state is given by
\beq
|G\rangle = \prod_{k=-\pi/2}^{\pi/2} c^\dagger_k |vac\rangle.
\eeq
It is known that the leading contribution to the entanglement entropy of an interval of length $L$ in this state is
\beq
S(L) = \frac{1}{3} \ln{(L/\epsilon)}
\eeq
with $\epsilon$ some cutoff length scale.  One way to obtain this result is to recast the tight binding model as a conformal field theory.  This is achieved by focusing on the two Fermi points at $k_F =\pm \pi /2$ and organizing the left and right moving states there in terms of a Dirac fermion in one dimension.  The Dirac fermion is a conformal field theory with $c=1$ and hence Eq. \ref{Sc} gives an entropy of $(1/3)\ln{(L/\epsilon)}$.  To be clear, this is the universal part of the entanglement controlled by the low energy degrees of freedom near the Fermi points.

Within the context of CFT, the $c$-theorem would imply that this entropy, i.e., $c$, must strictly decrease under the RG flow.  For example, by imposing Lorentz invariance entanglement monotonicity tells us that we cannot start with one Dirac point, add a relevant perturbation, and end up with two Dirac points.  However, we now construct an RG flow from a scale invariant but not conformally invariant theory to the Dirac fermion in which the analog of $c$, i.e. the coefficient of the log term in $S$, does not decrease.

The construction is simple.  In terms of the tight binding model above, we adjust the $w_\ell$ so that the Fermi points remain at $k=\pm \pi/2$ but the dispersion has vanishing first derivative there.  This is achieved by setting $w_{\ell} = 0$ for $\ell$ even, and by tuning $w_3$ relative to $w_1$.  The resulting system has $\partial_k \epsilon_k(k = \pm \pi/2_) =0$ and $\partial_k^2 \epsilon (k = \pm \pi/2) = 0$ (by symmetry).  The first non-vanishing derivative is thus the third and so the energy of excitations near the Fermi points scale like
\beq
\delta \epsilon \sim (k-k_F)^3.
\eeq
This is a scale invariant system with dynamical exponent $z=3$, e.g., energy is related to momentum as $\omega \sim q^z$.  Nevertheless, the ground state of this system is manifestly identical to that of the $z=1$ system above (with only $w_1$ non-zero).  Since the ground is the same, the entanglement entropy of the $z=3$ system is also the same as the entanglement entropy of the $z=1$ system.  A straightforward RG flow exists between the two systems provided we introduce a small Fermi velocity.  In terms of $v_F = \partial_k \epsilon_k$ and $v''_F = \partial_k^3 \epsilon_k$ we have a momentum scale $\Lambda^2 = v_F/v''_F$.  At time scales shorter than $(v''_F \Lambda^3)^{-1}$ the physics is controlled by the $z=3$ fixed point while at long times we recover the $z=1$ physics.  Nevertheless, we have shown there is no change in the entanglement entropy along this flow and hence a strict monotonicity result fails in terms of the coefficient of the logarithm in $S$.

We can also produce many generalizations of this result, for example, we can construct flows from any odd $z$ to any other odd $z'$ with $z' < z$.  Furthermore, we can arrange for the analog of $c$ to actually increase by further modifying the dispersion.  Consider a dispersion $\epsilon_k$ which near the chemical potential has the form
\beq
\epsilon_k - \mu = v_3 (k-k_F)^3 - v_1 (k-k_F) + ...
\eeq
with $v_1$ small.  This model is similar to the one we considered above, but with a negative velocity $v_1$.  At high energy the $v_1$ term has no appreciable effect and we have effectively the $z=3$ model consider above.  However, as we probe the system at lower energies and longer wavelengths the single zero crossing at $k=k_F$ effectively splits into three Fermi points at $k=k_F$ and $k = k_F \pm \sqrt{v_1/v_3}$.  Thus the low energy theory has $c=3$ and $z=1$ while the original fixed point has $c=1$ and $z=3$ where by $c$ we mean the coefficient of $\ln{(L)}$ in the entropy.

Having exhibited a variety of entanglement monotonicity violations for fermions in one dimension, we now turn to another simple problem of bosons of interest where a completely unentangled state flows to an entangled state under a relevant deformation.

\section{Boson superfluid}

Consider a bosonic lattice model on a $d$ dimensional hypercubic lattice.  We have bosonic operators $b_r$ on each site satisfying the usual commutation relations $[b_r , b_{r'}^\dagger] = \delta_{rr'}$.  The Hamiltonian is taken to be
\beq
H_b = - w \sum_{<rr'>} b_r^\dagger b_{r'} - \mu \sum_{r} n_r + U \sum_r n_r (n_r -1)
\eeq
where $n_r = b_r^\dagger b_r$ and the hopping term is a sum over nearest neighbors.  We consider the ground states of $H_b$ as a function of $\mu$ assuming $w \gg U$.  When $\mu <0$ the ground state is simply the vacuum satisfying
\beq
b_r | G(\mu <0)\rangle = b_r |vac\rangle =0.
\eeq
On the other hand, when $\mu > 0$ the ground state is a superfluid with a non-zero density of bosons.  The point $\mu=0$ is a critical point separating the vacuum and superfluid phases of the system.

Indeed, the $\mu=0$ point is actually scale invariant with dynamical exponent $z=2$, but the ground state remains the trivial vacuum state.  Furthermore, while the presence of the interaction modifies the scattering properties of multi-particle states, the ground state contains no particles and hence remains the trivial unentangled vacuum state.  A simple field theory for this critical point can be obtained in terms of a coarse-grained Bose field $\psi(x,t)$ with action
\beq
\mathcal{S} = \int d^d x dt \left( \psi^\dagger i \partial_t \psi - \frac{1}{2m} |\nabla \psi|^2 + \mu |\psi|^2 - g |\psi|^4\right).
\eeq
The scaling symmetry is easily visible when $g=0$: we send $x \rightarrow \lambda x$, $t \rightarrow \lambda^2 t$, and $\psi \rightarrow \lambda^{-d/2} \psi$ which leaves $\mathcal{S}$ invariant if $\mu =0$.  Furthermore, note that the interaction is actually irrelevant when $2d > d+z = d+2$ or $d> 2$.  More generally, the model has a controlled fixed point with $z=2$ at non-zero $g$ in $d=1$ and $d=2$ dimensions (see Ref. \cite{subir} for a review), but as discussed above, the ground state remains the trivial vacuum state in all cases.

When $\mu > 0$ we obtain the superfluid phase which has a linearly dispersing sound mode associated with the spontaneous breaking of the boson number symmetry.  Neglecting the physics of the zero mode (which can be included as needed, see Ref. \cite{2011arXiv1112.5166M}), we have the physics of a single relativistic free scalar field.  The details of this field theory or of the more complete lattice description need not concern us.  All we need to know is the ground state of the model is entangled with the entanglement entropy of a ball well described by Eqs. \ref{Sc}, \ref{SF}, and \ref{Sa} in $d=1,2,3$ dimensions.

Now again we have a simple RG flow from the $z=2$ system to the $z=1$ superfluid phase.  Starting from the UV fixed point with $z=2$ we add a chemical potential $\mu >0$ which is a relevant operator at the $z=2$ critical point.  The RG flow then takes us to the $z=1$ superfluid phase.  The physics of this RG flow is straightforward.  The system has a finite density $n$ of bosons, and at distances short compared to $n^{-1/d}$ the system looks empty ($z=2$ fixed point) while at long distances the system looks like a conventional superfluid ($z=1$ fixed point). Thus we have a simple and controlled RG flow from a $z=2$ system with an unentangled ground state to a $z=1$ system with an entangled ground state.  This trivial result has important implications.  First, the total entropy is clearly increasing along the RG flow.  Second, because the $z=2$ theory is unentangled, there is no non-trivial entanglement quantity we can construct from it that will decrease under RG flow (because it has no entanglement).

We now return to fermions to construct even more serious violations of entanglement monotonicity.  We make heavy use of the special features of fermions at finite density to obtain our results.

\section{Fermi surface}

Consider now a free Dirac fermion in any dimension with action
\beq
\mathcal{S}_\psi = \int d^d x dt \bar{\psi} i \gamma^\mu \partial_\mu \psi,
\eeq
and where $\{\gamma^\mu,\gamma^\nu\} = 2 g^{\mu \nu}$ with $g^{\mu \nu}$ the flat spacetime metric.  This system is a conformal field theory in any dimension, but we may easily perturb it with a Lorentz breaking perturbation to drive it into a new gapless phase.  The action has a symmetry under $\psi \rightarrow e^{i \theta} \psi$ which leads to a $U(1)$ conserved current $J^\mu = \bar{\psi} \gamma^\mu \psi$.  Using this current, we may add a chemical potential to the relativistic action to obtain the new action
\beq
\mathcal{S}'_\psi = \int d^d x dt \left[\bar{\psi} i \gamma^\mu \partial_\mu \psi + \mu J^0 \right].
\eeq
This chemical potential is a relevant perturbation at the free Dirac fixed point and drives a renormalization group flow into a new phase of matter.  See also the recent work of Ref. \cite{dirac_perchempot} for another simple example using Dirac fermions and a periodic chemical potential.

The ground state of this new system is a free Fermi gas with a Fermi surface.  The dispersion of the relativistic fermion is $\epsilon_k = |k|$, and the ground state of $\mathcal{S}'$ is obtained by filling up energy levels until $\epsilon_{|k| = k_F} = \mu$ (assume $\mu > 0 $).  Thus we have $\mu = k_F$ in units where the Fermi velocity is one.  Now the crucial question becomes, how does the entanglement of a $d$-ball scale in this new system?  In fact, not only is the new system $\mathcal{S}'$ not a conformal field theory, it doesn't even obey the area law for entanglement.  Using by now well-known methods, we can show that the entanglement entropy of a $d$-ball of radius $L$ scales like
\beq
S(L) \sim (k_F L)^{d-1} \ln{(k_F L)}
\eeq
with a known prefactor \cite{ee_f1,ee_f2,bgs_f1,2006PhRvB..74g3103L,2007JMP....48j2110F,2009arXiv0906.4946H}.  Except for $d=1$ this entanglement entropy does not even have the same scaling form as the CFT result, hence no direct comparison is possible.  Still, it must be admitted that the Fermi surface system is more entangled, e.g., has faster scaling of entanglement entropy with $L$, by any reasonable definition of that term.  Thus we have again a trivial violation of entanglement monotonicity.

To obtain a precise result that encodes this violation, consider the following modification of the above story.  Suppose we also add some irrelevant interactions to the Dirac fixed point in addition to the chemical potential.  These interactions, which are by assumption irrelevant at the Dirac fixed point, are constructed so that they lead to a non-zero marginally irrelevant interaction at the Fermi surface fixed point.  In detail, we assume that the interactions generate an attractive interaction in a superconducting pairing channel, the so-called BCS channel.  To be concrete, consider $d=2$ spatial dimensions with a circular Fermi surface.  Then the different BCS channels are indexed by an angular momentum $\ell = 0,1,2,...$, and the coupling $V_\ell$ obeys the RG equation
\beq
\frac{d V_\ell}{ds} = -\frac{V_\ell^2}{4\pi}
\eeq
where $s$ is the RG time, i.e. the momentum cutoff is $\Lambda(s) = \Lambda_0 e^{-s}$ \cite{RevModPhys.66.129}.  We arrange for one of the $V_\ell$ with large $\ell$ to be negative in which case the coupling grows even more negative under the RG flow.  Thus pairing and superconductivity will eventually set in the chosen angular momentum channel.  If we further arrange the interactions so that the favored gap structure is $\Delta_k \propto \langle c_k c_{-k} \rangle \propto \sin{\ell \theta_k}$ with $\theta_k = \tan^{-1}(k_y/k_x)$, then we obtain as a ground state a superconductor with $2 \ell$ nodes along the Fermi surface, e.g., the dispersion is
\beq
\sqrt{\epsilon_k^2 + \Delta_k^2}
\eeq
which has zeros whenever both $\epsilon_k$ and $\Delta_k$ vanish, e.g., on the Fermi surface and when $\ell \theta_k = n \pi$.  The low energy theory is then $2\ell$ Majorana nodes with linear but anistropic dispersion.  Indeed, near such a node the dispersion relation reads
\beq
\sqrt{v_F^2 k_\parallel^2 + v_\Delta k_\perp^2}
\eeq
The entanglement entropy of the resulting theory, in the case when $v_F = v_\Delta$ is simply that of $2 \ell$ relativistic Majorana fermions.  In particular, it has the form of Eq. \ref{SF} with $F \propto \ell$ and hence arbitrarily large compared to the UV $F = F_{Dirac}$.  Our construction achieves this violation by passing through an intermediate non-relativistic finite density state (the Fermi surface) before spontaneously breaking the global $U(1)$ symmetry to return to a scale invariant with $\propto \ell$ massless fermions.

Another interesting point raised by the construction is the role of anistropy.  Indeed, the scenario we have considered here has a similar low energy structure of that of the cuprate superconductors with a $d$-wave gap ($\ell=2$).  In those systems, $v_F \neq v_\Delta$, and the entanglement entropy of a disk is not given by the usual isotropic formula.  In fact, we can easily recover an isotropic system by rescaling one of the lengths, however this will distort the disk and convert it into an ellipse.  Crucially, the the entanglement entropy of an ellipse depends on the precise shape of the ellipse and the universal term is not known to decrease under the RG flow.  Furthermore, in the cuprate case the directions corresponding to $k_\perp$ and $k_\parallel$ are different for different nodes.  Hence we cannot do a single rescaling to simultaneously bring all the nodes to an isotropic point.  Thus such anisotropic but linearly dispersing systems may also provide an even simpler obstacle to generalizing the Lorentz invariant entanglement monotonicity theorems.

Of course, the system we have constructed is heavily tuned, breaks rotational invariance, breaks $U(1)$ invariance, and is generally a somewhat complicated physical system.  Even the so-called Majorana nodes do not sit at zero momentum as would be conventionally assumed in high energy physics.  Nevertheless, the basic low energy physics is indeed that of relativistic massless Majorana fermions.  Thus we do have an RG flow from a relativistic system (free Dirac fermion) to another relativistic system (many Majorana fermions) via an intermediate non-relativistic phase, or otherwise one must show how the subtle features of the superconducting state invalidate the comparison.  In any event, the simple fact that the free Dirac fermion flows to the Fermi gas, a much more highly entangled state, under relevant deformation by a chemical potential implies that entanglement is rather non-monotonic under the RG.

\section{Majorization}

We now shift focus slightly and ask a related question.  In the context of Lorentz invariant systems where the monotonicity theorems do hold, it is very interesting to ask if there is any more general construction from which these theorems follow.  For example, why only $d$-balls and why only the von Neumann entropy?  The Renyi entropy, defined by
\beq
S_n(A) = \frac{1}{1-n} \ln{\left(\text{tr}(\rho_A^n)\right)},
\eeq
is another candidate for monotonic quantity.  However, unlike the entanglement entropy, which obeys the strong subadditivity bound
\beq
S(AB) + S(BC) \geq S(ABC) + S(B),
\eeq
the Renyi entropy is not strong subadditive in general.  Hence there is something very special about the entanglement entropy. Nevertheless, perhaps the Renyi entropy is still monotonic.

As an example, in one spatial dimension the Renyi entropy of a single interval is a simple modification of Eq. \ref{Sc} given by
\beq
S_n = \frac{c}{6} \left(1+ \frac{1}{n}\right) \ln{(L/\epsilon)}.
\eeq
Since all these entropies are controlled by the same quantity, the central charge $c$, it follows that since $c$ is monotonic, all the Renyi entropies are also.  However, note that here we only mean that the UV and IR coefficients satisfy an inequality, not that the Renyi entropy is a monotonic function along the RG flow.  We must also be careful with the cutoff since as $n\rightarrow 0$ the Renyi entropy measures the total support of the density matrix which depends on the regulator.

On the other hand, we know in higher dimensions, three spatial dimensions for example, that non-spherical regions involve other anomaly coefficients besides $a$ which are known not to decrease under RG.  Hence in general only the entanglement of a $3$-ball obeys a monotonicity theorem.  Similarly, we know that Renyi entropies do not obey strong subadditivity, so at the very least the monotonicity proof in Ref. \cite{2012PhRvD..85l5016C} breaks down.  Of course, this does not mean that the Renyi entropy of a $d$-ball is not monotonic, indeed it monotonic, in the sense described above, when $d=1$.  However, we will now demonstrate that it cannot in general be monotonic in $d>1$.

To show this, it is useful to make a brief detour to discuss the concept of majorization.  Consider two normalized density matrices $\rho_1$ and $\rho_2$ for a fixed quantum system $\mathcal{H}$ ($\text{dim}(\mathcal{H}) = \chi$) with their eigenvalues $p_{i \alpha}$ ($i=1,2$, $\alpha=1,...,\chi$) in decreasing order.  We say that $p_1$ majorizes $p_2$ ($p_1 \succ p_2$) if
\beq
\sum_{\alpha=1}^k p_{1\alpha} \geq \sum_{\alpha=1}^k p_{2 \alpha}
\eeq
for all $k$.  Furthermore, a function $f$ from $\mathbb{R}^{\chi} \rightarrow \mathbb{R}$ is called Schur convex (concave) if $p \succ q$ implies that $f(p) \geq f(q)$ ($f(p) \leq f(q)$) (see Ref. \cite{PhysRevLett.83.436} for a discussion).  Majorization is a useful tool in the context of entanglement entropy in part because all Renyi entropies are Schur concave functions.  As a simple example, the maximally mixed state with probabilities $q^\star$ given by
\beq
q^\star_\alpha = \frac{1}{\chi}
\eeq
is majorized by every other probability distribution.  Hence the Schur concavity of $S_n$ implies that $S_n(p) \leq S_n(q^\star)$ for all $p$, and since $S_n(q^\star) = \ln{(\chi)}$ we obtain the basic inequality that $S_n(p) \leq \ln{(\chi)}$.

Remarkably, since in one dimensional CFTs all Renyi entropies of a single interval are controlled just by $c$ and since the Renyi entropies fully determine the spectrum of the corresponding density matrix, it follows that a majorization relationship is obeyed.  In detail, the spectrum of the density matrix at the IR fixed point, $q_{IR}$, and the spectrum of the density matrix at the UV fixed point, $q_{UV}$, obey $q_{UV} \prec q_{IR}$.  This relationship has been checked in a variety of cases both in the context of field theory and also in various lattice models realizing CFTs in their low energy physics \cite{PhysRevLett.90.227902,PhysRevA.71.052327}.

Here we use the recent results of Ref. \cite{2013arXiv1304.6402S} to show that such a majorization condition cannot hold for $d$-balls when $d>1$.  There it was shown that for $n\rightarrow 0$, $S_n$ is controlled by the thermal entropy density of the corresponding CFT.  Since the coefficient of the thermal entropy density is not monotonic along RG flows, it follows that the Renyi entropy is not monotonic for sufficiently small $n$, and hence that the spectra of the $d$-ball density matrices do not obey a majorization relationship in general.

To illustrate this non-monotonicity, we give a simple example involving an $O(N)$ invariant field $\vec{\phi}$.  Let us work in $D=2+1$ spacetime dimensions in imaginary time.  The action at the Gaussian fixed point is
\beq
\mathcal{S}_N = \frac{1}{2} \int d^D x (\partial \vec{\phi} )^2.
\eeq
The thermal energy per unit volume at temperature $T$ (in units where the speed is one) is
\beq
\epsilon(T) = N \int \frac{d^2 k}{(2\pi)^2} \frac{|k|}{e^{|k|/T}-1}.
\eeq
This expression evaluates to
\beq
\epsilon(T) = N \frac{T^3}{2\pi} \int \int_0^\infty dx \frac{x^2}{e^x - 1} = N \frac{T^3}{2\pi} \zeta(3) \Gamma(3).
\eeq
The entropy density is then
\beq
s_{Gaussian} = \frac{3}{2} N \frac{T^2}{2\pi} \zeta(3) \Gamma(3).
\eeq

On the other hand, we can also consider the Wilson-Fisher fixed point.  The Gaussian fixed point is unstable to the Wilson-Fisher fixed point and flows to it when a certain combination of the relevant operators $\vec{\phi}^2$ and $\vec{\phi}^4$ are added to the action.  We may access the Wilson-Fisher fixed point in the large $N$ limit as discussed in Ref. \cite{subir}.  The effective mass due to interactions scales with temperature as $m(T) = 2\ln{\left(\frac{1+\sqrt{5}}{2}\right)} T$.  The energy density is
\beq
\epsilon(T) = N \int \frac{d^2 k}{(2\pi)^2} \frac{\sqrt{k^2+m^2}}{e^{\sqrt{k^2+m^2}/T}-1},
\eeq
and upon switching to the variable $q^2 = k^2+m^2$ and doing the angular integral we find
\beq
\epsilon(T) = \frac{N}{2\pi} \int^\infty_{m} dq q \frac{q}{e^{q/T} -1}.
\eeq
Finally, we may write
\beq
\epsilon = \frac{N T^3}{2\pi } \int_{x_0}^\infty dx \frac{x^2}{e^x -1},
\eeq
which is almost the energy of the Gaussian theory.  Taking the ratio we find
\beq
\frac{\epsilon_{WF}(T)}{\epsilon_G(T)}= 1 - \delta
\eeq
with
\beq
\delta = \frac{\int_0^{x_0} dx \frac{x^2}{e^x -1}}{\int_0^{\infty} dx \frac{x^2}{e^x -1}},
\eeq
and where $x_0 = \sqrt{(1+\sqrt{5})/2}$.  We also find that
\beq
\frac{S_{WF}}{S_G} = 1-\delta.
\eeq

Now because $\delta$ is greater than zero and non-vanishing in the large $N$ limit, the flow from the Gaussian fixed point to the Wilson-Fisher fixed point is actually monotonic as regards the thermal entropy.  However, we can use the above the results to construct a non-monotonic flow by considering the flow from the Wilson-Fisher fixed point to the fixed point describing the symmetry broken phase.  This phase has $N-1$ Goldstone bosons and (modulo some subtleties about the zero mode) has the same thermodynamics and entanglement as $N-1$ massless scalar fields.  Thus the thermal entropy density actually increase along the RG flow because, if $s_0$ is the entropy of a single massless scalar, we have $N s_0 (1-\delta) < (N-1) s_0$ for sufficiently large $N$.  Hence because the thermal entropy density controls the Renyi entropy of a disk at small Renyi index, it follows that the Renyi entropy is also not monotonic and hence no majorization condition can be obeyed.

\section{Discussion}

In this paper we have investigated the structure of entanglement along RG flows in a wide variety of systems.  We have exhibited RG flows from unentangled to entangled states, from theories with different dynamics but identical entanglement, and from a Lorentz invariant fixed point to another Lorentz-invariant fixed point via a non-Lorentz invariant intermediate Fermi gas state.  We also showed that a majorization condition is not obeyed by the eigenvalues of the reduced density matrix in more than one dimension.  The general picture is then two fold: first, the varieties of entanglement scaling are greatly multiplied when Lorentz invariance is lost, and second, entanglement does not generally decrease under RG flows.

There are still many interesting questions to answer about the interplay between entanglement and renormalization even in Lorentz invariant systems.  One example is provided by topological phases in three dimensions.  In two dimensions, topological entanglement entropy contributes to $F$ despite the fact that it arises from a gapped system where as we normally associated $F$ with gapless degrees of freedom.  However, in three dimensions the analog of topological entanglement entropy does not contribute to $a$, so it would be interesting to know if there is a generalization of $a$ that includes some topological contribution and is still monotonic.  We also do not yet know of entanglement monotones in dimensions greater than three.

Our results imply that with no additional restrictions, no quantity defined purely from the wavefunction will be universally monotonic under RG flows.  However, this leaves open the possibility that by restricting to certain symmetries, by considering families of states, or by studying dynamical properties or perturbations, a kind of entanglement monotonicity result might still be found.  It will be interesting to study in much more detail the structure of entanglement along RG flows and to attempt to formulate generalized monotonicity principles.  Indeed, such results could have a profound impact on our understanding of quantum matter, for example, by providing additional evidence for the intuition that Nature prefers ground states of low entanglement (which is itself a special case of the general intuition that useful entanglement is fragile).

\textit{Acknowledgements.}
I thank T. Grover and S. Sachdev for useful discussions about this work.  I am supported by a Simons Fellowship through Harvard University.

\bibliography{RGmonoCM}

\end{document}